# Multiple Solutions in Fluid Mechanics

Lun-Shin Yao
Department of Mechanical Engineering
Arizona State University
Tempe, Arizona 85213

## Abstract

The principle of multiple solutions of the Navier-Stokes equations discussed in this paper is not directed at any particular problems in fluid dynamics, nor at any specific applications. The non-uniqueness principle states that the Reynolds number, above its critical value, is insufficient to uniquely determine a flow field for a given geometry, or for similar geometries. It is a generic principle for all fluid flow and its transportation properties, but is not well understood. It compliments the current popular bifurcation theories by the fact that multiple solutions can exist on each stable bifurcation branch.

## Introduction

Stokes (1851) was the first to recognize that a single dimensionless parameter, Re, uniquely determines a flow field for a given geometry. Later work by Reynolds (1883) on the onset of turbulence in the flow through tubes led to this being termed the *Reynolds number*. Since then, the concept of dynamic similarity has been firmly established and has governed the development of fluid dynamics for more than a century. For example, a laminar fully-developed tube flow is a function of Re only, and is independent of the upstream (or entry) condition. Similarly, for an equilibrium turbulent tube flow or turbulent boundary layer, the Reynolds number uniquely determines their mean flows and statistical properties. The upstream flow conditions can only affect the flow transient, which will be quickly forgotten as they reach their equilibrium states. This is the concept of uniqueness in fluid dynamics that we have been taught, and is the rule, which guides the progress in fluid dynamics. A comprehensive review of more recent work in multiple solutions can be found in Gelfgat & Bar-Yoseph (2004).

In this paper, we state a different fact, contrary to the conventional wisdom, that the Reynolds number above its critical value is *insufficient* to determine a flow field uniquely. The non-uniqueness of fluid flow means that flows of different wavenumbers or frequencies can exist on each stable *bifurcation* branch. The upstream or ambient conditions can persist and determine the final equilibrium wavenumber or frequency whose accessible band is slightly narrower than the linear instability boundary. Consequently, the non-uniqueness principle differs from the recent popular bifurcation theories. Some experiments have revealed this fact, but we constantly treat them as isolated incidents due to our firm belief in the principle of the dynamic similarity. This may be due to the lack of a theory to put those experimental evidences together to show that the Reynolds number plus the *environments* determine a flow field uniquely.

Since Coles (1965) studied that Taylor-Couette flows have multiple solutions for a fixed Reynolds number, extensive studies have shown complex flow structures for Taylor-Couette flows for various geometric configurations. Similar phenomena have been identified for Benard instabilities induced by temperature gradient or surface tension. These are both closed flow systems. The observations consistently indicate that the initial condition determines the final supercritical equilibrium state for a given Reynolds number. In general, we have considered that the multiple solutions for these two types of closed flows are special cases. The concept of dynamic similarity is intact for all other flows. We will review this group of experiments first. The review is limited to the concept of non-uniqueness without touching any detailed physics of the flow, since the Taylor-Couette flow is probably the most studied flow in the history of fluid dynamics.

Even though the idea of coherent structure for large-scale turbulence has been around for a long time (Dryden 1948, Townsend 1956), flow visualization of turbulent mixing layers by Brown and Roshko (1974) was the first convincing demonstration of its existence. Their picture of the turbulent mixing layer was the key to the gate of current research in turbulence. They collected data on spreading rates for plane mixing layers and found a large scatter about one hundred percent in the measured values. Subsequent research has found that the evolution of mixing layers depends on the nature of the environmental noise, which is facility and location dependent. In our terminology, multiple solutions exist and depend on the upstream conditions.

Surprisingly, no communication has ever existed between the group who study turbulent mixing layers and those for Taylor-Couette flows, even though the members of the Cal Tech Group originated both experiments. Now, it is clear that initial conditions determine the equilibrium state for a closed flow system, and upstream conditions determine that for an open flow system. One is a temporal problem and the other, a spatial problem. Only one fundamental principle, nonlinear wave interaction, governs the evolution of both systems. We will review the aspect of non-uniqueness of turbulent mixing layers and other relevant flows.

The physics associated with the non-uniqueness in fluid dynamics can be described by the Navier-Stokes equations and will not change due to different methods of solution. We introduce a Fourier-eigenfunction spectral method below for two reasons. First, for a certain class of problems, the required CPU time for our method is only a small fraction of that for a more powerful Fourier-Chebyshev spectral method. Second, two equations of historical importance, the Landau and Ginzburg-Landau equations, are the limiting cases of our formulation. In 1945, Landau intuitively proposed the nonlinear form of the finite amplitude equation for the instability disturbance. Stuart and Watson (1960) were the first to rationally derive the Landau equation. The equation has dominated the development of nonlinear instability of fluid dynamics for more than thirty years. Almost all modern instability theories are built upon the basis of the Landau equation. Many important progresses leading to the present understanding of fluid dynamics would be impossible without the Landau equation. The simple fact that chaotic solutions of fluid dynamics were first found from the Ginzburg-Landau equation underscores its importance. In the following, we will show that the nonlinear terms in both equations are self-interaction of the dominant wave. Neither equation considers the possible nonlinear interaction among all waves. Some research has included more than one amplitude equation to study the nonlinear interaction between a few waves. Unfortunately, the number of the interacting waves considered was too few and incomplete to reveal the importance of the non-uniqueness principle in fluid dynamics.

The nonlinear interactions among Taylor vortices with different wavenumbers was studied by Yao & Ghosh Moulic using a weakly nonlinear theory (1995a) and nonlinear theory (1994b) formulated with a continuous spectrum. They represented the disturbance by a Fourier integral and derived an integrodifferential equation for the evolution of the amplitude density function of a continuous spectrum. Their formulation allows nonlinear energy transfer among all participating waves and remedies the shortcomings of equations of the Landau-Stuart type. Numerical integrations of this integrodifferential equation indicate that the equilibrium state of the flow depends on the wavenumber and amplitude of the initial disturbance as observed experimentally and cannot be determined uniquely on each stable bifurcation branch. The accessible wavenumbers of the nonlinear stable range lie within the linear unstable range minus the range of the side-band instability. This agrees with Snyder's observation (1969). These results also show that sideband instability is a consequence of nonlinear energy transfer among different interacting waves as well as with the mean flow. *The trend of energy transfer is determined by the stability characteristics of the mean flow.*

The analysis of mixed convection in a vertical annulus using a nonlinear theory with continuous spectrum (Yao and Ghosh Moulic 1994a, 1995b) yields the identical conclusion as that from the study of Taylor-Couette flows. They treated the spatial problem as a temporal problem. This is possible by following a control mass system. Plots of the evolution of the kinetic-energy spectra for the two problems are identical in shape and differ only in amplitudes and ranges of wavenumbers. A typical plot of the evolution of kinetic energy is shown in Figure 1. This indicates that the evolution of energy spectra provides universal and fundamental information for flows. The eigenfunctions, which are associated with the stability characteristics of the mean flow, are different for different flows. The selection of the equilibrium wavenumber is due to nonlinear wave interactions. The *selection* principles outlined from the numerical results are quite universal and listed below:

1) When the initial disturbance consists of a single dominant wave within the nonlinear stable region, the initial wave remains dominant in the final equilibrium state. Consequently, for a slowly starting flow, the critical wave is likely to be dominant.

2) When the initial condition consists of two waves with finite amplitudes in the nonlinear stable region, the final dominant wave is the one with the higher initial amplitude. If the two waves have the same initial finite amplitude, the dominant wave seems to be the one closer to the critical wave. On the other hand, if the initial amplitudes are very small, the faster growing wave becomes dominant.

3) When the initial disturbance has a uniform broadband spectrum, the final dominant wave is the fastest linearly growing wave, if the initial amplitude is small. On the other hand, if the uniform noise level is not small, the critical wave is the dominant equilibrium one.

4) Any initial disturbance outside the accessible frequency range will excite its Subharmonics or superharmonics whichever are inside the accessible frequency range. The accessible frequency range is the linear stability range minus the range of the side-band instability.

**Experiments for Closed Flows**

Taylor (1923) demonstrated theoretically and experimentally that circular Couette flow becomes unstable when the speed of the inner cylinder is increased beyond a certain critical value. His experiments showed that the instability leads to a new steady axisymmetric secondary flow in the form of regularly spaced vortices in the axial direction. The documentation that Re is insufficient to uniquely determine a flow field in Taylor double cylinder experiment went back a half century ago. Pai (1943) noted that the flow in his apparatus consists of either 4 or 6 Taylor vortices depending on the past history. Using hot-wire anemometry Pai showed that the velocity field differs for two possible cases. Even though he worked at very large Re and his work is primarily qualitative, it is the first experiment to show that Re does not determine turbulence uniquely. Hagerty (1946) observed that the wavelength can increase or decrease by a factor of nearly two if the length of the fluid column changes. The aspect ratios of both Pai's and Hagerty's apparatus are very small that end effects dominates the results.

Coles (1965) observed that the axisymmetric Taylor-vortex flow becomes unstable as the angular speed of the inner cylinder is increased further, beyond a second critical value. This instability results in a wavy-vortex flow, with azimuthally propagating waves superposed on the Taylor vortices. Coles found that the spatial structure of the wavy-vortex flow, characterized by the axial and azimuthal wavenumbers, is not a unique function of the Reynolds number and boundary conditions. Different equilibrium states could be achieved at the same Reynolds number by approaching the final Reynolds number with different acceleration rates, and by rotating and then stopping the outer cylinder. Coles recognized that time-dependent amplitude equations of Landau-Stuart type fail to predict and explain the non-uniqueness of Taylor-Couette flows and stated, "In view of the important advances in nonlinear technique which are contained in this analytical work by Stuart, Davey, Segel, and others, it is disappointing that such analyses do not serve the present need; but this is unfortunately the case."

The non-uniqueness of the equilibrium state observed by Coles (1965) was subsequently observed in time-independent Taylor-vortex flow by Snyder (1969), Burkhalter & Koschmieder (1974) and Benjamin (1978). Snyder (1969) found that, while the wavelength at the onset of instability was unique, Taylor vortex flows with different wavenumbers could be obtained at the same value of the Reynolds number by varying the initial conditions. He observed that there was *a band of accessible wavenumbers*, smaller than the band that can grow according to linear instability theory. The region between the linear instability boundary and the band of the accessible wavenumbers is now known as sideband instability. Burkhalter & Koschmieder (1974) found that the range of axial wavelengths for stable Taylor-vortex flow is quite large. Benjamin (1978) observed different spatial states even in an annulus so short that only three or four vortices could be accommodated. Fenstermacher, Swinney & Gollub (1979) studied the transition to turbulence in Taylor-Couette flow using laser-Doppler anemometry. They also found that the different spatial states had different spectra and transition Reynolds numbers. Presumably, these different final states at the same value of the Reynolds number are functions of the initial conditions.

Multiple solutions of complex structures have also been found for Benard instabilities (Chen and Whitehead 1968, Koschmieder 1993).

**Experiments for Open flows**

Before we review turbulent flows, we would like to discuss an experiment of a laminar flow in curved channels. Mees (1994) periodically modulated the flow rate by very low rms amplitudes of only a small fraction of one percent of the mean flow rate. He tried a wide range of forcing frequencies and found the accessible equilibrium frequencies are bounded in a finite range. This agrees with the non-uniqueness principle of Coles' observation for Taylor-Couette flows and of the analytical solution for mixed convection. Mees showed that any initial disturbance outside the accessible frequency range will excite its subharmonics or superharmonics whichever are inside the accessible frequency range. This is the fourth selection rule. A plot of the averaged energy frequency spectra looks very much like Figure 1. The flow visualization for curved-duct flows differs, of course, completely from those for Taylor-Couette flows or mixed convection in a vertical annulus. This is because their flow patterns depend on their particular stability eigenfunctions.

Ho and Huang (1982) forced a two-dimensional mixing layer by modulating the inlet flow rates of two upstream chambers at rms amplitudes ranging from 0.02 percent to 0.09 percent of the mean speed. The perturbation is homogeneous across the span of the mixing layer. The Reynolds numbers in the downstream test region lie in the range 100-1000. The response frequency measured near the trailing edge of the splitting plate shows that the accessible frequencies are bounded in a finite range (see their Figure 2). The forcing frequency, lower than the accessible range, excites its superharmonics. This provides the initial condition for the mixing layer.

One common argument concerning the turbulent mixing layer is that the large-scale structure is instability, may be a transient turbulence, and is not equilibrium turbulence. One important contribution by Ho and Huang, which has not been widely appreciated, is their measurement of the turbulent boundary layers on the splitting flat plate. Flow transition on a flat plate occurs usually fast as does its reaching the equilibrium state. Their Figure 2 clearly indicates that the frequency of the energy-containing eddy of a turbulent boundary layer on a flat plate depends on the upstream conditions and can be any frequency within the accessible frequency band. This is hard evidence leading to the conclusion that turbulent shear flows have multiple solutions.

The characteristics of the mixing layer differ from those of the boundary layer on the splitting plate. The mixing layer is unstable to a long wave modulation. Its accessible frequency range is lower than that for a boundary layer on a plate. Consequently, the measured original frequency immediately downstream of the splitting plate falls outside of the accessible frequency range of the mixing layer. Therefore, the amplitude of the original response spectrum declines and excites its Subharmonics, which increase downstream as the flow approaches its equilibrium state. A plot of amplification of the energy content (see Figures 15, 16, 18 and 19 of Ho & Huang, 1982) looks very similar to the evolution of kinetic energy shown in Figure 1. The shifting of accessible frequency ranges causes the merging of several vortices simultaneously. They showed that the merging of vortices is the mechanism for increasing the spreading rate of the mixing layer.

In terms of nonlinear theory, the merging of vortices is driven by sub-harmonic resonance. An examination of their data clearly shows that "pairing" of vortices follows exactly the same selection rules outlined above for nonlinear wave interactions. Thus, "pairing" is determined by the stability characteristics of the mean flow and not by vortex dynamics.

Similar experiments intended to study the dynamic process of turbulent mixing layers have been conducted with a mechanical flap at the trailing edge (Oster & Wygnanski 1982; Weisbrot & Wygnanski 1988). They found that the oscillations of the flap have no significant effect on the initial velocity distribution nor on the total turbulent energy near the trailing of the splitting plate. Nevertheless, the forced oscillations caused remarkable differences in the mixing layer downstream. Oster and Wygnanski found that, in the absent of forcing, the energy spectrum is at 230 hz which is a subharmonic of the shedding frequency near the trailing edge of the flap (see their Figure 2). With increasing amplitude of the surging, the energy content at the forcing frequency (60 hz) increases markedly. This follows exactly the second selection rule outlined above. The frequency spectra (their Figures 32 and 34) show that the modulation of the mean flow is of the same order as the amplitude of the dominant wave. This agrees with the nonlinear analyses of Taylor-Couette flow or mixed convection.

Weisbrot and Wygnanski noted that fluctuations, locked in phase with the disturbance frequency, are responsible for the initial growth of the mixing layer. It has been demonstrated by the nonlinear theory that an equilibrium state can only be achieved if the disturbance waves are locked in phase. Their data show that the wave associated with the forcing frequency persists very far downstream and that the spectra of the broadband fluctuations associated with small-scale turbulence depend on the large-scale structure (see their Figures 9-14). Similar phenomena have been observed in two-dimensional, small-deficit, turbulent wakes (Wygnanski, Champagne & Marasli 1986). This shows that the memory of initial conditions can persist far downstream into the final period of decay after the large-scale structure has disappeared. This effect is commonly referred to as *history* effect of turbulence. The accessible range of equilibrium frequencies of the large-scale structure is determined by the nonlinear energy transfer and differs for different mean flows. This turbulent history effect is selective. Because each experimental facility has its own characteristic free-stream disturbances, the waves generated in a particular facility may be unique to that facility only. Furthermore, the characteristic free-stream disturbance may change for different locations and for different seasons. The history effect of turbulence found in laboratories is exactly the non-uniqueness in Coles' sense. Since the ambient conditions for a real flow are not a controllable design factor, a non-unique solution introduces uncertainty to real flows.

**Nonlinear Theory**

**<u>Fourier-Eigenfunction Spectral Formulation</u>**

A fully-developed mean flow in cylindrical polar coordinates is used as an example to illustrate the nonlinear theory. This limitation of fully-developed mean flows can be removed with a substantial increase in the required computational time. The formulation is for a temporal flow development, but can be easily converted to a spatial flow development.

The velocity components, (u, v, w) and temperature, θ, are decomposed into mean values plus disturbances. It is not necessary that the magnitude of the disturbances is smaller than that of the mean values. They are

$$\mathbf{u} = [u, v, w, \theta] = [u', v', W_0(r) + w', \Theta_0(r) + \theta'], \tag{1}$$

where $W_O$ and $\Theta_O$ are the mean axial velocity and temperature. The disturbance is expressed as

$$\mathbf{u}(\mathbf{r},t) = \mathbf{U}(\mathbf{r}) + \int_{-\infty}^{\infty} \sum_{n=-\infty}^{\infty} \sum_{m=1}^{\infty} A_m(k,n,t)\tilde{\mathbf{u}}_m(k,n,r)e^{i(kz+n\phi)}dk, \qquad (2)$$

where $A_m$ is the amplitude density function, and $\tilde{\mathbf{u}}_m$'s are the linear stability eigenfunctions. Substituting (2) into the Navier-Stokes equations and projecting along the direction of the adjoint eigenfunctions result in

$$\frac{\partial A_m}{\partial t} + i\omega_m A_m = \sum_{m_1=1}^{\infty} \sum_{m_2=1}^{\infty} \sum_{n_1=-\infty}^{\infty} I\left(k,n,m,m_1,m_2,n_1,t\right), \qquad (3)$$

where

$$I = \int_{-\infty}^{\infty} b\left(k_1,k-k_1,n_1,n-n_1,m_1,m_2,m\right)A_{m_1}\left(k_1,n_1,t\right)A_{m_2}\left(k-k_1,n-n_1,t\right)dk_1,$$

and the b's are constants of wave interactions. The linear terms of the Navier-Stokes equations represent the mean-flow convection of the disturbances, the distortion of the disturbances by the mean-flow stresses and the body forces, and diffusion. In the generalized coordinates of eigenfunctions, they are reduced to a single term, which determines the growth or decay of the wave due to linear energy transfer. For a fixed wavenumber, $\omega_m^i$ is more negative for larger m. This indicates that eigenfunctions for larger m have more dissipation capacity, or it can be interpreted as a smaller eddy. But, this eddy is not in a spherical shape and is elongated. This introduces anisotropy into the flow structure. For laminar flows, the feedback of small-scale motions to large-scale structures is negligible. For turbulent flows, the magnitude of small-scale motions may not be much smaller than that of large-scale structures. Their structures are determined by the stability characteristics of the mean flow.

The eigenfunction expansion (2) has reduced the continuity, momentum, and energy equations to the system of integro-differential equations for the amplitude density functions without any approximations. Thus, the solution of equations (3) represents an exact solution of the Navier-Stokes equations. It is worth pointing out that one of the major difficulties in the numerical solution of the incompressible Navier-Stokes system is the simultaneous enforcement of the no-slip boundary conditions and the incompressibility constraint. Since the vector of basis functions used in the expansion (2) are solutions of the linearized Navier-Stokes equations, they individually satisfy the incompressibility constraint as well as the boundary conditions. Therefore, the expansion (2) automatically satisfies the boundary conditions and the continuity equation. Thus, the numerical solution of the system of equations (3) is much simpler than the numerical solution of the Navier-Stokes equations. It may also be noted that straightforward evaluation of the convolution product representing the nonlinear terms in equation (3) is inefficient if the number of terms in the truncated eigenfunction expansion used in the numerical solution is large. However, pseudo-spectral evaluation of the convolution product can make the numerical solution of the equations (3) a viable efficient alternative to the numerical solution of the Navier-Stokes equations. A preliminary study shows that the required CPU time can be as little as one sixth of that needed to solve the Navier-Stokes equations by a Fourier-Chebyshev collocation spectral method.

## **Weakly Nonlinear Theory**

In the following, we will show that classical weakly nonlinear instability theories are special limiting cases of (3). The weakly nonlinear development of the disturbances may be studied by expanding the Fourier amplitudes in a perturbation series. The maximum amplification rate predicted by linear instability theory is used as the expansion parameter $\varepsilon$. The amplification rate predicted by linear theory for the m-th eigenmode of the wave with axial wavenumber $k$ and azimuthal wavenumber $n$ may be expressed in terms of $\varepsilon$ as $\omega_m^I(k,n) = \varepsilon\, a_m(k,n)$ where $\omega_m^R$ is the (real) frequency. This makes $a_m(k,n)$ a constant of order one. The amplitude density function is expanded in a perturbation series as

$$\tilde{A}_m(k,n,t) = \varepsilon A_{m,1}(k,n,t,T_1,T_2) + \varepsilon^2 A_{m,2}(k,n,t,T_1,T_2) + \cdots,\qquad(4)$$

where $T_1 = \varepsilon t$ and $T_2 = \varepsilon^2 t$ are slow time scales. Substitution of the expansion (4) into equation (3) results in a set of equations

$$\frac{\partial A_{m,l}}{\partial T} = a_m \, A_{m,l} \; + \sum_{m_1, m_2} \; \sum_{n_1} \; \widetilde{I}_3^S \left(k, n, m, m_1, m_2, n_1, T\right)$$

$$+ \sum_{\substack{m_1, m_2, \\ m_3, m_4}} \; \sum_{n_1, n_2} \; \widetilde{I}_4^S \left(k, n, m, m_1, m_2, m_3, m_4, n_1, n_2, T\right) , \tag{5}$$

where $T = \varepsilon\, t$ and

$$\widetilde{I}_3 = \int_{-\infty}^{\infty} \left[b + \varepsilon \widetilde{b}\right] A_{m_1,l} \left(k_1, n_1, T\right) \, A_{m_2,l} \left(k - k_1, n - n_1, T\right) e^{i\Omega_{3w}t} \delta \left(\Omega_{3w}\right) dk_1 \tag{6}$$

and

$$\widetilde{I}_4 = \int_{-\alpha}^{\infty}\!\!\int \varepsilon \widetilde{c} \; A_{m_1,l} \left(k_1, n_1, T\right) A_{m_3,l} \left(k_2, n_2, T\right)$$

$$A_{m_4,l} \left(k - k_1 - k_2, n - n_1 - n_2, T\right) e^{i\Omega_{4w}t} \delta \left(\Omega_{4w}\right) dk_1 dk_2 \; . \tag{7}$$

Equations (6) and (7) are resonant triads and quartets. The resonance conditions are

$$\Omega_{3w} = \omega_m^R(k, n) - \omega_{m1}^R(k_1, n_1) - \omega_{m2}^R(k - k_1, n - n_1) = 0 , \tag{8}$$

and

$$\Omega_{4w} = \omega_m^R(k, n) - \omega_{m1}^R(k_1, n_1) - \omega_{m2}^R(k_2, n_2)$$

$$- \omega_{m3}^R(k - k_1 - k_2, n - n_1 - n_2) = 0 , \tag{9}$$

respectively.

## Landau Equation

The equation (5) for the evolution of the amplitude density function of a continuous spectrum contains as a special case the equation describing the evolution of the amplitude of a discrete monochromatic wave. The amplitude density function for a discrete wave with axial wavenumber $k_0$ and azimuthal wavenumber $n_0$ may be expressed in the form

$$A_{m,1}\left(k,n,t\right) = \left[\; A_0\left(t\right)\delta\left(k-k_0\right)\delta_{n,n_0} + A_0^*\left(t\right)\delta\left(k+k_0\right)\delta_{n,-n_0}\; \right]\delta_{m,1}\;, \tag{10}$$

where $\delta(k)$ represents the Dirac delta function, $\delta_{i,j}$ represents the Kronecker delta, and the asterisk denotes complex conjugates. We have retained only the least stable mode ($m = 1$) in the leading-order amplitude density function expressed in equation (10). Substitution of (10) into (5) leads to

$$\frac{dA_0}{dt} = a_0 A_0 + a_1\left|A_0\right|^2 A_0\,, \tag{11}$$

where $a_0 = \omega_1^1(k_0,n_0)$ is the amplification rate predicted by linear stability theory for the least stable eigenmode of the wavenumber $(k_0,\,n_0)$ and

$$a_1 = \hat{c}\left(k_0,-k_0,k_0,n_0,-n_0,n_0\right) + \hat{c}\left(-k_0,k_0,k_0,-n_0,n_0,n_0\right)$$

$$+\hat{c}\left(k_0,k_0,-k_0,n_0,n_0,-n_0\right) \tag{12}$$

is the second Landau constant. Equation (11) is the Landau equation describing the temporal evolution of the amplitude of a discrete monochromatic wave. For a monochromatic wave, no three waves can satisfy the resonance condition; the second Landau constant is contributed by the self-interaction of the wave, which forms resonance quartets. The nonlinear term in the Landau equation does not represent energy transfer with other waves. The equation (5) for the evolution of the amplitude density function of a continuous spectrum can also be reduced to a set of N ordinary differential equations describing the evolution of the amplitude of N discrete waves. Then, the set of equations includes the weakly nonlinear energy transfer among all waves.

## **Ginzburg-Landau Equation**

It is worth noting that although equation (5) describes the temporal evolution of the amplitude density function in wave space, the integral formulation does include spatial variations of the disturbances through the Fourier integral transform (2), and is not restricted to periodic disturbances. In the special case of a slowly varying wave packet with an azimuthal wavenumber $n_0$, and a spectrum confined to a small neighborhood of bandwidth $\delta$ around the minimum critical wavenumber $k_c$, equation (5) becomes

$$w'\left(r,\phi,z,t\right) = \tilde{A}\left(k_c,n_0,z,t\right)\tilde{w}_1\left(k_c,n_0,r\right)e^{i\left[k_c z + n_0\phi - \omega_1^R(k_c,n_0)t\right]} + \text{c.c.} \qquad (13)$$

where

$$\tilde{A}\left(k_c,n_0,z,t\right) = \int_{k_c-\delta}^{k_c+\delta} A_1\left(k,n_0,t\right)e^{i(k-k_c)z+i\left[\omega_1^R(k_c,n_0)-\omega_1^R(k,n_0)\right]t}\,dk$$

$$= \varepsilon \int_{-\delta/\varepsilon}^{\delta/\varepsilon} A_1\left(k_c+\varepsilon K,n_0,t\right)e^{i\varepsilon Kz+i\left[\omega_1^R(k_c,n_0)-\omega_1^R(k_c+\varepsilon K,n_0)\right]t}\,dK \qquad (14)$$

is the slowly-varying envelope of the wave train, and c.c. denotes the complex conjugate. As in equation (10), we have retained only the least stable mode (m = 1) in equation (14). In order to derive an equation for the envelope of the wave packet, we multiply equation (5) by

$$\delta_{n,n_0}\exp\left\{i\left[\omega_1^R(k_c,n_0)-\omega_1^R(k,n_0)\right]t\right\}$$

$$\delta_{m,1}\exp\left\{i\left(k-k_c\right)z\right\}$$ and integrate with respect to k from $k_c$-$\delta$ to $k_c$+$\delta$. We expand the linear amplification rates and frequencies in a Taylor series around k = $k_c$ :

$$a(k,n_0) = a(k_c,n_0) + \frac{1}{2}\frac{d^2a}{dk^2}(k_c,n_0)(k-k_c)^2 + \dots,$$

and $\qquad\qquad\qquad\qquad\qquad\qquad\qquad\qquad\qquad\qquad\qquad\qquad\qquad (15)$

$$\omega^R(k, n_0) = \omega^R(k_c, n_0) + \frac{d\omega^R}{dk}(k_c, n_0)(k - k_c) + \ldots$$

It may be noted that at the minimum critical wavenumber $k_c$, $\frac{da}{dk}(k_c, n_0) = 0$. With (14) and , in the limit as $\varepsilon \to 0$, it can be shown that equation (5) reduces to the equation describing the evolution of a wave packet and is known as Ginzburg-Landau equation:

$$\left( \frac{\partial}{\partial t} + c_g \frac{\partial}{\partial z} \right) \tilde{A} = a_0 \tilde{A} + a_2 \frac{\partial^2}{\partial z^2} \tilde{A} + a_1 \left| \tilde{A} \right|^2 \tilde{A}, \tag{16}$$

where $a_2 = -\frac{1}{2}\frac{d^2 a}{dk^2}(k_c, n_0)$ and $c_g = \frac{d\omega^R}{dk}(k_c, n_0)$ is the group velocity. The nonlinear term of (16) is similar to the one in the Landau equation and represents the self-interaction of the carrying wave. Thus, the Ginzburg-Landau equation cannot be used for nonlinear interaction of different waves.

### **Indirect Nonlinear Wave interaction**

Both the numerical results and experimental data indicate that energy is drawn out from the mean flow to support the disturbance wave and its harmonics. The amplitude of mean-flow modulation is as large as that of the dominant wave. If there is no energy transfer to waves other than the selected disturbance wave, a set of simpler amplitude equations can be derived to replace equation (5) of a continuous spectrum. Rigorously, the expansion should include the amplitude density functions for the mean flow, the dominant disturbance wave and its harmonics. We assume that the harmonics are negligibly small to simplify the algebra in order to demonstrate the concept clearly. Substitution of the amplitude density function,

$$A(k, t) = \left[ A_0\, \delta(k) + A_0^*\, \delta(k) \right] + \left[ A_{k_0}\, \delta(k - k_0) + A_{k_0}^*\, \delta(k + k_0) \right], \tag{17}$$

into equation (5) results in

$$\frac{dA_{k_0}}{dt} = a(k_0)A_{k_0} + \left[a_{31}A_0 A_{k_0}\right] + \left\{\left[a_{41}\left|A_{k_0}\right|^2 + a_{42}\left|A_0\right|^2\right]\right\}A_{k_0},\tag{18}$$

and

$$\frac{dA_0}{dt} = a(0)A_0 + \left\{\left[b_{31}\left|A_0\right|^2 + b_{32}\left|A_{k_0}\right|^2\right]\right\} + \left\{\left[b_{41}\left|A_{k_0}\right|^2 + b_{42}\left|A_0\right|^2\right]\right\}A_0,\tag{19}$$

where $a_{ij}$ and $b_{ij}$ are interacting constants. $A_0$ is the amplitude function for the deformation of the mean flow, and $A_{k_0}$ is for the disturbance wave. The terms inside of the first square bracket on the right-hand side of (18) and (19) are from a three-wave resonance. The terms in the second square bracket are from a four-wave resonance. The terms multiplied by $a_{31}$ and $a_{42}$ represent the nonlinear energy transfer between the mean flow and the disturbance wave, and are not included in the Landau and Ginzburg-Landau equations. The terms represent the *indirect* wave interaction and have been overlooked by most weakly nonlinear theories. Any waves, which do not satisfy the resonance condition, can still interact indirectly by interaction with the mean flow. This fact has been verified by the numerical solution of (3). The term multiplied by $a_{41}$ represents the self-interaction of the disturbance wave and is the only nonlinear term in the Landau and Ginzburg-Landau equations. It is clear that only linear energy transfer with the mean flow, the terms multiplied by the growth rate, *a*, is considered in the Landau and Ginzburg-Landau equations. Equations similar to (18) and (19) have been derived for the interaction of two disturbance waves, but none of them have considered an additional equation for the deformation of the mean flow. They have ignored the fact that the mean flow acts like a medium to provide energy to sustain the disturbances and assumed that the deformation of the mean flow is of smaller order. Without the terms of nonlinear energy transfer between the mean flow and the disturbance wave, the energy is not properly accounted for.

The Figure 6.23 of Mees (1994) shows the importance of the *indirect* mode, a direct nonlinear energy transfer between the mean flow and all waves. The indirect mode is responsible for the rising of a broadband spectrum near the inlet. The broadband spectrum decays and transfers energy to the dominant wave and its harmonics as they move downstream. The resonance conditions cannot always be met among waves. This agrees with our numerical solution of (3). The indirect mode is also likely to be the mechanism for flow transition to turbulence when the nonlinear energy transfer balances with the linear dissipation.

## **Numerical Examples**

The numerical solution of equation (5) requires the evaluation of two integrals. The first integral involves a quadratic nonlinearity, while the second involves a cubic nonlinearity. The solution of equation (3), on the other hand, requires the evaluation of only one integral involving a quadratic nonlinearity. For the same accuracy the weakly nonlinear theories require, at least, double the storage and CPU time than a direct solution of the nonlinear formulation (3). Moreover, equation (3) is equivalent to the Navier-Stokes equations. Thus, from a computational point of view, it is preferable to solve equations (3) directly. Furthermore, (3) can be effectively solved simultaneously by parallel machines.

For the Taylor-Couette problem for $\eta = r_1 / r_2 = 0.5$ at $Re = 88.1$ slightly above its critical state, $Re = 68.1$, the range of wavenumbers permitted for supercritical Taylor vortices according to linear theory is $1.6 \le k \le 5.6$. The results indicate that the equilibrium state depends on the initial condition and is not unique. The range of equilibrium wavenumbers was found to be narrower than the span of the neutral curve from linear theory. Flows with wavenumbers outside this range, but within the unstable region of linear theory are found to be unstable and to decay, but to excite another wave inside the narrow band. This result is in agreement with the Eckhaus and Benjamin-Feir sideband instability. The results also show that linearly stable long and short waves can also excite a wave inside this narrow band through nonlinear wave interaction. The results suggest that the selection of the equilibrium wavenumber is due to a nonlinear energy transfer process, which is sensitive to initial conditions.

In the following we will discuss the results of one case for the initial disturbance at $k_i = 3$. The solution of (3) agrees with the direct numerical solution of the Navier-Stokes equations by a Fourier-Chebyshev spectral method up to the fourth decimal place. This is not a surprise because both solutions are exact for the problem. The equilibrium results of (3) are included on

Table 1. The equilibrium amplitude at $k = 3$ agrees with the results of (5) and slightly differs from that of (11) due to the indirect mode. On the other hand, the largest modification of the mean flow is due to the third eigenfunction, $m = 3$. This shows the insufficiency of the classical theories (11) and (16) which only retain the first eigenmode, $m = 1$. The classical theories for monochromatic waves can only be considered qualitatively acceptable even very close to the neutral stability curve.

New physics is indeed revealed by our computation. Each superharmonic represents a smaller vortex of less strength. The Taylor vortex is composed of a sequence of vortices whose axial wavelengths are integer fractions of the dominant one. Also, a Taylor vortex is not a pure stationary wave, but with a small contribution from standing waves due to the eigenmodes as highlighted on Table 2. The standing wave causes a small oscillation of the amplitude of the Taylor vortices. In experiments, the dominant Taylor vortex has been observed to fluctuate with respect to time. It is difficult to visualize the fine structures in a laboratory. This shows that our computation can provide more detailed flow structures than experimental measurements. This demonstrates that the advantage of the new formulation (3) is not only that it requires much less CPU time than other numerical methods; it can also reveal more detailed flow structures.

The equilibrium states of mixed convection in a vertical annulus for $\eta = 0.375$, Re = 100, Ra = 200 and Pr = 0.6 are not unique either. The linear stability analysis indicates that the basic flow is unstable to disturbances within a narrow wavenumber band between 0.23 and 1.13. The selection of the final equilibrium wave for mixed convection follows the same principles as those for Taylor-Couette vortices. The evolution of unstable waves for the initial disturbance $k_i = 0.25$ is plotted in Figure 1. The final dominant wave is $k = 0.5$ due to the sideband instability. It is worthwhile to note that the mean-flow distortion is much larger than the amplitude of the dominant wave. This shows that the assumption of the classical theories for monochromatic wave that the modification of the mean flow is a small-order effect is not valid.

Another important implication that the equilibrium state of the *mean* flow and the wave components are not unique is that time-averaged turbulent mean flows are not unique for a given Reynolds number. Thus, the values of time-averaged turbulent statistical quantities do not equal the ensemble average for stationary turbulence. From an application point of view, only the time average has physical significance.

## Conclusion

I hope I have convinced you that non-uniqueness is a generic property for all fluid flow. Reynolds number alone is insufficient to uniquely determine a flow field and its transport properties. Low-amplitude environmental perturbations can have profound effects on the determination of the equilibrium state. For problems near the onset of instability, the dynamic similarity ensured by the Reynolds number requires only a minimum modification. The variation of engineering data, such as Nusselt numbers and flow resistance, can be substantial, but they involve no exciting new physics. For turbulence, I believe, modification will not be small.

TABLE 1.  TAYLOR-COUETTE FOR $K_I$ = 3 AND RE=88.1

AMPLITUDES OF THE DIFFERENT EIGENMODES

| m | k = 0 | k = 3 | k = 6 | k = 9 | k = 12 |
|---|-------|-------|-------|-------|--------|
| 1 | 0.364E-01 | 0.865E-01 | 0.177E-01 | 0.322E-02 | 0.825E-03 |
| 2 | 0.195E-11 | 0.871E-02 | 0.114E-02 | 0.107E-03 | 0.198E-03 |
| 3 | 0.458E-01 | 0.447E-02 | 0.577E-02 | 0.834E-03 | 0.343E-04 |
| 4 | 0.300E-11 | **0.449E-02** | **0.512E-02** | 0.586E-03 | 0.557E-04 |
| 5 | 0.127E-02 | **0.449E-02** | **0.512E-02** | 0.438E-03 | 0.933E-04 |
| 6 | 0.107E-11 | 0.532E-03 | **0.113E-02** | **0.278E-03** | **0.753E-04** |
| 7 | 0.141E-02 | 0.503E-04 | **0.113E-02** | **0.278E-03** | **0.753E-04** |

TABLE 2.  FREQUENCIES OF THE DIFFERENT EIGENMODES

| m | k = 0 | k = 3 | k = 6 | k = 9 | k = 12 |
|---|-------|-------|-------|-------|--------|
| 1 | 0.216E-13 | 0.303E-09 | 0.606E-09 | 0.910E-09 | 0.121E-08 |
| 2 | 0.216E-13 | 0.303E-09 | 0.606E-09 | 0.910E-09 | 0.121E-08 |
| 3 | 0.216E-13 | 0.303E-09 | 0.606E-09 | 0.910E-09 | 0.121E-08 |
| 4 | 0.216E-13 | **0.920E+00** | **-0.855E+00** | 0.910E-09 | 0.121E-08 |
| 5 | 0.216E-13 | **-0.920E+00** | **0.855E+00** | 0.910E-09 | 0.121E-08 |
| 6 | 0.216E-13 | 0.303E-09 | **-0.108E+01** | **-0.108E+01** | **0.207E+01** |
| 7 | 0.216E-13 | 0.303E-09 | **0.108E+01** | **0.108E+01** | **-0.207E+01** |
|   |   |   |   |   |   |

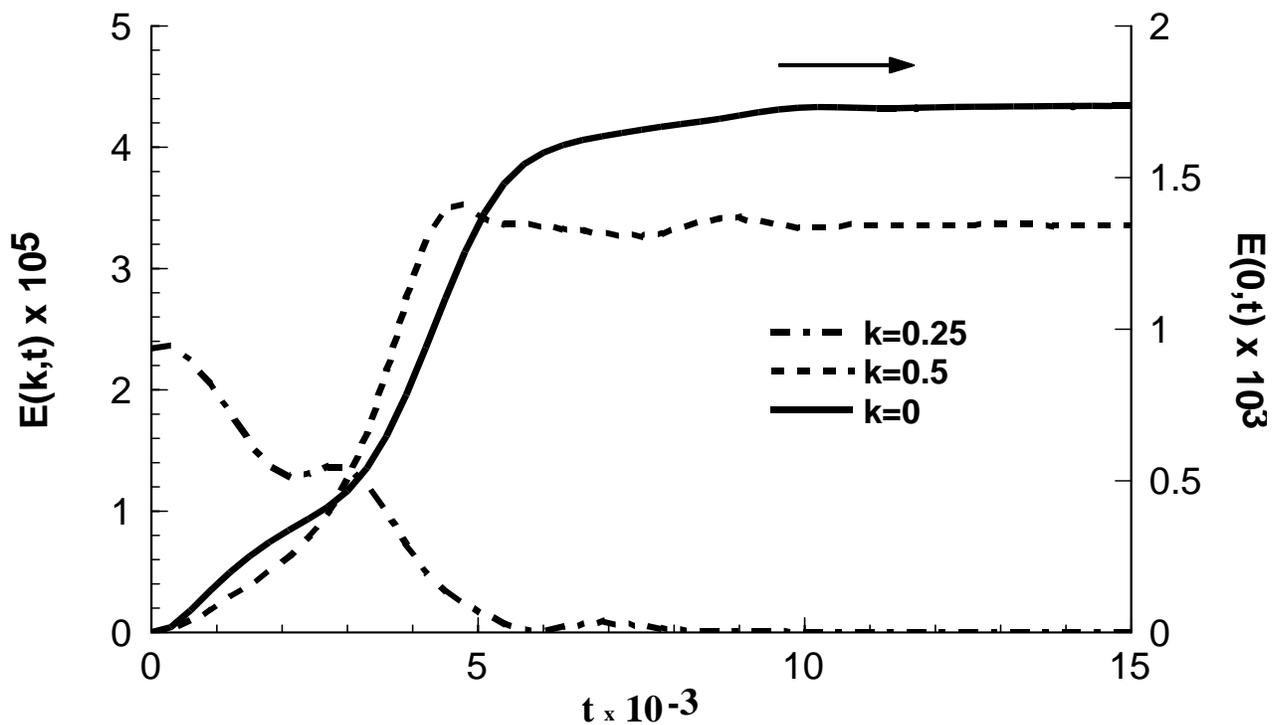

**FIGURE 1. EVOLUTION OF KINETIC ENERGY OF THE DOMINANT WAVE WITH THE INITIAL DISTURBANCE AT K=0.25.**